\documentclass[reprint,superscriptaddress,showpacs,preprintnumbers,nofootinbib,amsmath,amssymb,aps,onecolumn,prd,
floatfix
]{revtex4}

\usepackage{graphicx}% Include figure files
\usepackage{epsfig}
\usepackage{dcolumn}% Align table columns on decimal point
\usepackage{bm}% bold math
\usepackage{amssymb}
\usepackage{amsmath}
\usepackage{subfigure}
\usepackage{cancel}
\usepackage[colorlinks]{hyperref}
\usepackage[usenames,dvipsnames]{color}% add hypertext capabilities
\hypersetup{
     breaklinks=true,
    pdfstartview={FitH},    % fits the width of the page to the window
    colorlinks=true,       % false: boxed links; true: colored links
    linkcolor=blue,          % color of internal links
    citecolor=red,        % color of links to bibliography
    filecolor=magenta,      % color of file links
    urlcolor=blue,           % color of external links
    anchorcolor=green,      % Color for anchor text
    linktocpage=true
}

\newcommand{\Mpl}{M_{\textrm{Pl}}}

\def\doi{http://doi.org}
\def\arxiv{http://arxiv.org/abs}

\begin{document}

\title{Baryogenesis in the paradigm of quintessential inflation}
\author{Safia Ahmad}
\email{safia@ctp-jamia.res.in}
\affiliation{Centre for Theoretical Physics, Jamia Millia
Islamia,
New Delhi-110025, India}
\author{Antonio De Felice}
\email{antonio.defelice@yukawa.kyoto-u.ac.jp  }
\affiliation{Yukawa Institute for Theoretical Physics, Kyoto University, Kyoto 606-8502, Japan}
\author{Nur Jaman}
\email{nurjaman@ctp-jamia.res.in} 
\affiliation{Centre
for Theoretical Physics, Jamia Millia Islamia, New Delhi-110025,
India}
\author{Sachiko Kuroyanagi }
\email{skuro@nagoya-u.jp }
\affiliation{Department of Physics, Nagoya University, Chikusa, Nagoya 464-8602, Japan }
\affiliation{Instituto de F\'isica Te\'orica UAM-CSIC, Universidad Auton\'oma de Madrid, \\Cantoblanco, 28049 Madrid, Spain}
\author{M.~Sami}
\email{msami@jmi.ac.in}  \affiliation{Centre
for Theoretical Physics, Jamia Millia Islamia, New Delhi-110025,
India}\affiliation{Maulana Azad National Urdu
University, Gachibowli, Hyderabad-500032,India}\affiliation{Institute for Advanced Physics and Mathematics, Zhejiang University of Technology,\\ Hangzhou, 310032, China}
\affiliation{ Eurasian  International Center for Theoretical Physics, Eurasian National University, Nur-Sultan
010008, Kazakhstan.}
\begin{abstract}
We explore the possibility of baryogenesis in the framework of quintessential inflation. We focus on the model independent features of the underlying paradigm and demonstrate that the required baryon asymmetry can successfully be generated in this scenario. To this effect, we use the effective field theory framework with desired terms in the Lagrangian necessary to mimic baryon number violation \textit{\`{a} la} spontaneous baryogenesis which can successfully evade Sakharov's requirement allowing us to generate the observed baryon asymmetry in the equilibrium process. Our estimates are independent of the underlying physical process responsible for baryon number violation.
The  underlying  framework of quintessential inflation essentially includes the presence of kinetic regime after inflation which gives rise to blue spectrum
of gravitational wave background at high frequencies. In addition to baryogenesis, we discuss the prospects of detection of relic
gravitational wave background, in the future proposed
missions, sticking to model independent treatment. 
\end{abstract}

%\pacs{Valid PACS appear here}
\maketitle
\section{Introduction}
Standard model of universe has several remarkable successes to its credit including the synthesis of light elements in the early universe. However, the successful nucleosynthesis requires an important input, namely, the baryon to photon ratio, $\eta\equiv (b-\bar{b})/n_\gamma=(6.1\pm 0.3)\times 10^{-10} $ that remains constant from sufficiently early times till date. Since no anti-baryons are seen in the Universe, the asymmetry requires explanation and at present there is no satisfactory model for baryogenesis. One might think that the  asymmetry should have existed from the very beginning which, however, looks untenable from the point of view of inflation \cite{Guth:1980zm,starobinsky,Linde:1981mu} that would dilute any such miss match if it were there. It is therefore reasonable to think that there was non conservation of baryon number in the early universe which led to the observed asymmetry. Sakharov formulated three necessary conditions for baryogenesis: Baryon number non conservation, C and CP violation and the necessity of  out of equilibrium processes. The third condition is important as the thermal average of Baryon number operator vanishes identically in the equilibrium state in presence of CPT invariance. There have been efforts to generate the asymmetry using 
electro-weak scenario, however, the recent discovery of Higgs boson with mass around 125 GeV has depressed these efforts\footnote{The required asymmetry asks for the Higgs mass much less than the observed value.}. One is led to think that effective theory beyond the standard electroweak scenario might be exploited to capture the said effect. Secondly, one might try to avoid the third Sakharov condition \cite{Sakharov:1967dj} by noting that the non-generation of Baryon asymmetry in equilibrium uses CPT invariance and the fact that Baryon number operator is C and CP odd. Thus the ``no go"  can be evaded if CPT is violated and asymmetry can be generated in the equilibrium state \textit{\`{a} la} the spontaneous baryogenesis \cite{cohen1,Dolgov:1997qr,Trodden:1998ym,DeSimone:2016ofp,Takahashi:2003db,Arbuzova:2016qfh,Dasgupta:2018eha}.

The aforesaid can naturally be implemented in the paradigm of quintessential inflation \cite{Sami:2004ic,Hossain:2014zma,Peebles:1998qn,Spokoiny:1993kt,Peebles:1999fz,Peloso:1999dm,Copeland:2000hn,Dimopoulos:2000md,Majumdar:2001mm,Rosenfeld:2005mt,Sami:2004xk,Dimopoulos:2001ix,Giovannini:2003jw,Tsujikawa:2013fta,Hossain:2014xha,Dimopoulos:2017zvq,Linde,Ahmad:2017itq,Jaman:2018ucm,Hossain:2018pnf} where scalar field $\phi$ survives in the post inflationary era.
Quintessential inflation stands for a scheme of unification of inflation and late time acceleration using a single scalar field. In this case the underlying field potential is typically run away type
asking for an alternative mechanism of reheating. Steep potential in the post inflationary era is needed for the commencement of radiative regime. 
%And independence of field dynamics 
And independence of initial conditions of late time field evolution asks for a particular type of steepness of the potential. For instance, a steep exponential potential gives rise to scaling behavior such that the field energy density tracks the background. In this set up, it is  challenging to meet all the observational requirements related to inflation. It was recently shown that a generalized exponential potential \cite{Geng:2015fla,Geng:2017mic,Skugoreva:2019blk} can successfully describe inflation and despite its non standard form can give rise to scaling behavior in the post inflationary era.

The aforesaid constitutes essential features of the paradigm. However, since the scaling solution is non accelerating, one requires to exit from this regime to late time acceleration {\it \`a la} tracker \cite{Steinhardt:1999nw,Caldwell:1997ii,trac1,Chiba:2009gg}. Such a transition can, in particular, be realized by coupling the field to massive neutrino matter. The coupling gets generated dynamically only at late times as massive neutrino turn non- relativistic. The latter triggers minimum in the potential where the field can settle giving rise to late time acceleration\footnote{It is not know how to construct a field potential which successfully describes inflation and gives rise to tracker like evolution in the post inflationary era after recovery from the freezing regime without invoking an extra feature in the potential. 
It is, however, possible to build a single field potential giving rise to quintessential inflation with thawing realization; in this case,  the late time evolution is sensitive to initial conditions which is undesirable, see Ref. \cite{Dimopoulos:2017zvq} for details. }. As for reheating, one might try to use gravitational particle production\cite{ford} which is a universal mechanism caused by non adiabatic change in the space time geometry at the epoch when inflation ends. However, this mechanism is  inefficient such that the field remains in the kinetic regime for a long time before radiation domination could commence. During kinetic regime, the energy density in relic gravitational waves produced around the end of inflation enhances compared to the field energy density throwing  a challenge to nucleosynthesis constraint at the commencement of radiative regime. An alternative mechanism such as instant preheating might rescue the situation.

In the nutshell, in the framework of quintessential inflation, we have a scalar field at our disposal in the post inflationary era
which, apart from the aforesaid role assigned to it, should also facilitate spontaneous baryogenesis \cite{cohen1,DeFelice:2002ir}. To this effect, one might assume its coupling to a non-conserving  baryon current of the type, $\partial_\mu \phi J_B^\mu $ \cite{Dolgov:1997qr,DeSimone:2016ofp,Takahashi:2003db}.
In this case, the non conservation of baryonic current can be attributed to spontaneous violation of baryon symmetry. Such a term is obviously absent in the standard model of particle physics which, however, can be thought to be present in the effective theory with a cut-off to be fixed from observations. This is a particular way asymmetry could be generated in the equilibrium state.
In this paper, using the effective field theory framework, we explore the possibility of generation of baryon asymmetry, in the paradigm of quintessential inflation dubbed spontaneous baryogenesis adhering to model independent features of the scenario. We also discuss the prospects of detection of relic gravitational waves in the present set up in the forthcoming observational missions.

%%%%%%%%%%%%%%%%%%%%%%%%%%%%%%%%%%%%%%%%%%%%%%%%%%%%%%%%%%%%%%%%%%%
\section{Inflationary and Post inflationary dynamics}
\begin{figure}[ht]
\centering
\includegraphics[scale=0.75]{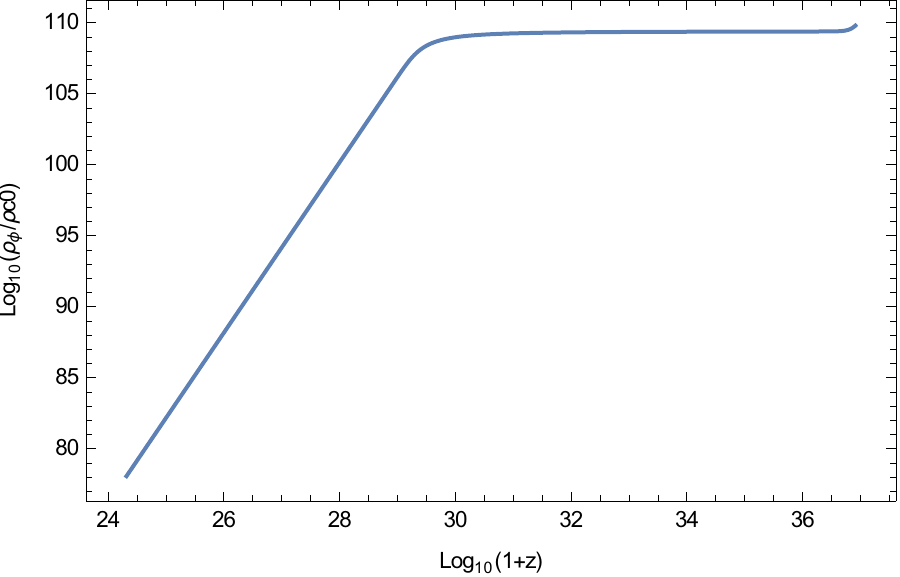}
 \caption{Figure shows the evolution of field energy density versus red-shift on log scale. It is clear from the plot that kinetic regime establishes fast after the end of inflation such that $H_{\rm end}\simeq H_{\rm kin}; T_{\rm end}\simeq T_{\rm kin}(a_{\rm end}\simeq a_{\rm kin})$, which we have used for the model independent estimates obtained in this paper. Obviously, the related numeric requires the knowledge of the underlying inflationary model; we have assumed the field potential to be a generalized exponential potential $V\sim \exp(-\lambda \phi^n/\Mpl^n)$ to obtain this plot. We also checked it in other models of quintessential inflation, for instance, the model discussed in Ref. \cite{Geng:2015fla}.}
\label{fig:rho}
\end{figure}

In this section, we bring out the basic features of the paradigm of quintessential inflation without resorting to a particular model. Irrespective of the underlying model, we can estimate the scale of inflation which is very close to $H_{\rm end}$ due to the small numerical value of the tensor-to-scalar ratio of perturbations $r$; the exact relation between them requires the knowledge of the underlying model$-$ the inflaton potential. Secondly, the commencement of kinetic regime takes place soon after inflation ends confirmed by numerical simulation using successful model of quintessential inflation, see Fig.\ \ref{fig:rho} and Ref.\ \cite{Ahmad:2017itq}.
In what follows, we present estimates on inflationary and post inflationary evolution required for the forthcoming analysis.

In the spatially flat Friedmann-Lemaitre-Robertson-Walker (FLRW) Universe, the metric is 
\begin{equation}
\label{metric}
ds^2 = -dt^2 + a(t)^2 \delta_{ij} dx^i dx^j,
\end{equation}
where $a(t)$ is the scale factor.
The action for the ordinary scalar field, $\phi$ minimally coupled to gravity is given by
\begin{equation}
\label{action}
S = \int d^4x \sqrt{-g} \left[ \frac{\Mpl^2}{2} R - \frac{1}{2} g^{\mu \nu} \partial_\mu \phi \partial_\nu \phi - V(\phi) \right],
\end{equation} 
where $V(\phi)$ is the potential of the scalar field. The variation of the action with respect to the metric $g^{\mu \nu}$ gives us the energy momentum tensor,
\begin{eqnarray}
T_{\mu \nu} = -\frac{2}{\sqrt{-g}} \frac{\delta S}{\delta g^{\mu \nu}} = \partial_\mu \phi \partial_\nu \phi -  g_{\mu \nu} \left[ \frac{1}{2} g^{\alpha \beta} \partial_\alpha \phi \partial_\beta \phi + V(\phi) \right],
\end{eqnarray} 
so that the energy density, $\rho_\phi$ and the pressure density, $p_\phi$ of the scalar field are given by
\begin{eqnarray}
&& \rho_\phi \equiv -T_0^0 = \frac{\dot{\phi}^2}{2} + V(\phi), \, \\
&& p_\phi \equiv T_i^i = \frac{\dot{\phi}^2}{2} - V(\phi),
\end{eqnarray}
and hence we have the following Einstein equations,
\begin{eqnarray}
\label{Friedmann_eqns}
\left(\frac{\dot{a}}{a} \right)^2 &=& \frac{1}{3 \Mpl^2} \rho_\phi , \\
\frac{\ddot{a}}{a} &=& -\frac{1}{6 \Mpl^2} (\rho_\phi+ 3p_\phi),
\end{eqnarray}
with $\dot{a}(t)/a(t)\equiv H$ being the Hubble parameter.
The variation of the action, Eq.~(\ref{action}), with respect to the field $\phi$ gives us the equation of motion of the field,
\begin{equation}
\label{eq:phi_evol}
\ddot{\phi} + 3 H \dot{\phi} + V'(\phi) = 0,
\end{equation}
where prime denotes the derivative with respect to the field $\phi$.
During inflation, the dominant component of the Universe is the inflaton field, $\phi$. So in order to have a phase of rapid expansion, we must have
\begin{eqnarray}
\rho_\phi + 3 p_\phi <0 ~~\Rightarrow ~~ \omega_\phi < -\frac{1}{3}\,,
\end{eqnarray}
where we have used the relation $p_\phi = \omega_\phi \rho_\phi$ with $\omega_\phi$ being the equation of state parameter of the inflaton field.

The energy scale associated with the inflation, $V_{\rm inf}$ {\it i.e.}\ the value of inflaton potential when the cosmological scales leave the horizon can be obtained by using the amplitude of scalar density perturbations \cite{Tsujikawa:2003zd},
\begin{equation}
A_s^2 = \frac{V_{\rm inf}}{150 \pi^2 \Mpl^4 \epsilon}\,,
\end{equation}
where $\epsilon$ is the slow-roll parameter given by
\begin{equation}
\epsilon \equiv -\frac{\dot{H }}{H^2} =\frac{\Mpl^2}{2} \left( \frac{V'}{V} \right)^2.
\end{equation}
So, from COBE normalization $A_s^2=4\times 10^{-10}$ \cite{cobe1},  we find that,
\begin{equation}
V_{\rm inf}^{1/4} = 0.014\times  r^{1/4} \Mpl \,,
\label{vinf}
\end{equation}
where $r$ is the tensor-to-scalar ratio and is related to the slow-roll parameter $\epsilon$ by the relation $r = 16 \epsilon$. Then the value of the Hubble parameter during inflation is
\begin{equation}
H_{\rm inf} = \frac{V_{\rm inf}^{1/2}}{\sqrt{3}M_{\rm Pl}}
=2.7\times10^{14}r^{1/2}{\rm GeV}.
\label{Hinfval}
\end{equation}
Inflation ends when the equation of state parameter $\omega_\phi$ takes the value, $\omega^{end}_\phi= -1/3$, which implies that
\begin{equation}
\dot{\phi}_{\rm end} = V_{\rm end}^{1/2}.
\label{phid}
\end{equation}
So the Hubble parameter at the end of inflation is
\begin{equation}
H_{\rm end} = \frac{V_{\rm end}^{1/2}}{\sqrt{2} \Mpl}\, .
\label{Hend2}
\end{equation}
Since the observation supports the small value of $r$ or $\epsilon$, it is a good approximation to take, $H_{\rm inf}\simeq H_{\rm end}$ though their exact relation asks for the underlying model.
Study of specific cases, reveals that the ratio, $H_{\rm inf}/H_{\rm end}$ is of the order of one; for instance in case of $V\sim \exp(-\lambda \phi^n/\Mpl^n)$, numerical results reveal that $ H_{\rm inf}/H_{\rm end}\simeq 1.8$ \cite{Ahmad:2017itq}. Hence,
throughout this paper, we assume $H_{\rm end}\simeq H_{\rm inf}$ which allows us to obtain the model independent estimates.

As shown in Fig.~\ref{fig:rho}, in a class of models of quintessential inflation, the scalar field enters the kinetic regime as soon as the inflation ends and the energy density evolves as,
\begin{eqnarray}
\rho_\phi = \rho_{\rm \phi,end} \left( \frac{a_{\rm end}}{a} \right)^6 \label{rho_phi1},
\end{eqnarray} 
where $\rho_{\phi,{\rm end}}=\frac{3}{2}V_{\rm end }$.  

Since the gravitational particle production is not efficient,
we implement instant preheating for the models of quintessential inflation \cite{Felder:1998vq}, which comfortably meets the requirement of nucleosynthesis constraint.  The radiation energy produced due to the mechanism of instant preheating is given by the following expression (see Refs.\ \cite{Sami:2004xk,Sami:2004ic} and the appendix for details)\footnote{In the framework of instant preheating, one assumes the coupling of inflaton with a scalar field $\chi$ which is also coupled to matter field.}
\begin{equation}
\rho_{\rm r} \approx \frac{g^2 V_{\rm end}}{8 \pi^3} \left( \frac{a_{\rm end}}{a} \right)^4.
\label{eq:rhor}
\end{equation}
Following inflation, Universe evolves in the kinetic regime until the radiation begins to dominate.
The scale factor at the commencement of radiative regime $a_r$ can be calculated by using $\rho_\phi(a_{\rm r})= \rho_{\rm r}(a_{\rm r})$, which gives us\footnote{Here we used the fact that at the end of inflation, $\omega^{\rm end}_\phi=-1/3 $ or $({\dot{\phi}}^2/2)_{\rm end}=V_{\rm end}/2$.}
\begin{equation}
\label{eq:arbaend}
\frac{a_{\rm r}}{a_{\rm end}} = \frac{\sqrt{12 \pi^3}}{g}.
\end{equation}
The number density of relativistic particles (radiation) produced by instant preheating is given by \cite{Hossain:2014zma} 
\begin{eqnarray}
n= \frac{g^{3/2} V_{\rm end}^{3/4}}{8 \pi^3}\left(\frac{a_{\rm end}}{a}\right)^3.
\end{eqnarray}
The thermal equilibrium is established when the Hubble parameter, $H \sim \Gamma = n \sigma$ with $\sigma$ being the annihilation cross section given by
\begin{eqnarray}
\sigma \sim \frac{\alpha^2 }{T_{\rm end}^2} \left(\frac{a}{a_{\rm end}}\right)^2,
\end{eqnarray}
where $\alpha$ is the coupling constant.
This leads to the scale factor $a_{\rm th}$ at which thermal equilibrium is established as\footnote{We assume that when the thermal equilibrium is established, the universe is dominated by scalar field in the kinetic regime which is reasonable as the field remains in the kinetic regime for long time after inflation ends in the paradigm of quintessential inflation.}
\begin{eqnarray}
\label{eq:thermala2nd}
\frac{a_{\rm th}}{a_{\rm end}} \sim \frac{2^{9/8}\pi^{3/2} T_{\rm end} }{\alpha~ g^{3/4} \Mpl^{3/4} H_{\rm end}^{1/4}}.
\end{eqnarray}

The radiation energy density is related to the temperature as 
\begin{equation}
\rho_{\rm r} = \frac{\pi^2}{30} g_* T^4,
\label{eq:rhordef}
\end{equation}
where $g_*$ is the total number of relativistic degrees of freedom which for $T \gtrsim 100$GeV is of the order of 100. 
Now, it is fair to assume that the radiation density follows the equilibrium distribution even before the equilibrium established \cite{Kolb:1990vq}. The temperature soon after the instant preheating $T_{\rm end }$ can be estimated using Eqs.~(\ref{vinf}), (\ref{eq:rhor}) and (\ref{eq:rhordef}), and is given by
\begin{eqnarray}
\label{eq:tend2}
T_{\rm end} \sim \frac{(15/2)^{1/4} g^{1/2}\Mpl^{1/2}H_{\rm end}^{1/2}}{\pi^{5/4} g_*^{1/4}}.
\end{eqnarray}
Using $H_{\rm end}\simeq H_{\rm inf}$, we find,
\begin{equation}
T_{\rm end} \simeq 1.3\times10^{-3} g^{1/2}r^{1/4}\Mpl\, .
\end{equation}
For the lower bound on $g$, see Eq.\ (\ref{eq:glimit}) and the tensor-to-scalar ratio of perturbations $r=0.05$, the estimated value of radiation temperature at the end of inflation is given by, $T_{\rm end} \approx 2.3 \times10^{13} ~\rm GeV $.
Assuming that $g_*$ does not change with temperature of the universe, which decreases as $T\propto a^{-1}$, and using Eq.\ \eqref{eq:thermala2nd}, we calculate the thermalization temperature,
\begin{eqnarray}
\label{eq:TthF}
T_{\rm th} \sim \frac{\alpha ~g^{3/4}  \Mpl^{3/4} H_{\rm end}^{1/4}}{2^{9/8}\pi ^{3/2}}.
\end{eqnarray}
Using, $H_{\rm end} \approx H_{inf}$, $\alpha=0.1$, the lower bound on $g$, see Eq.\ (\ref{eq:glimit}), and the upper bound on tensor-to-scalar ratio of perturbations, $r=0.05$, we find that, $T_{\rm th}\approx 2.7\times 10^{12} \rm GeV$. The temperature at the commencement of the radiative regime is given using Eqs.\ (\ref{eq:arbaend}) and (\ref{eq:tend2}) as
\begin{eqnarray}
\label{Tr}
T_{r} &=& T_{\rm end}\frac{a_{\rm end}}{a_r}
\sim \left(\frac{15}{288\pi^{11}g_*}\right)^{1/4}g^{3/2}M_{\rm Pl}^{1/2}H_{\rm end}^{1/2}.
\end{eqnarray} 
For generic values of $r$ and couplings used to estimate $T_{\rm th}$, we find that $T_r \approx 10^8\rm GeV$ such that $T_r/T_{\rm th} =a_{\rm th}/a_{r} \lesssim 10^4$.

The estimates presented above allow us to proceed further to implement baryogenesis in the framework under consideration. Since standard model of particle physics fails to comply with the observed value of baryon asymmetry, we shall use the effective field framework which has phenomenological character. We once again emphasize that the presented estimates are model independent and apply to the paradigm of quintessential inflation rather than its concrete realization.

%%%%%%%%%%%%%%%%%%%%%%%%%%%%%%%%%%%%%%%%%%%%%%%%%%%%%%%%%%%%%%%%%%%
\section{Quintessential Baryogenesis}
As mentioned in the introduction, inflaton field survives after inflation ends in the framework of quintessential inflation. Indeed, the scalar field $\phi$ evolves in the kinetic regime  for quite long before the commencement of radiation domination. One might imagine the interaction of $\phi$ with a non-conserved baryonic current which could be attributed to spontaneous symmetry breaking of some hypothetical U(1) symmetry \textit{\`{a} la} spontaneous baryogenesis.
The mentioned features are obviously absent in the standard model of particle physics. 
The additional structure in the Lagrangian would enter  with  a cut-off to be fixed from observation \textit{\`{a} la} the effective field construction.
It would, however, be necessary to check for the back reaction of the added terms in the Lagrangian on the cosmological dynamics of field $\phi$.
In the discussion to follow, we shall use the  mentioned framework to generate the required baryon asymmetry.
\subsection{General framework}
In view of the aforesaid, let us consider the effective Lagrangian density  of the following form \cite{DeFelice:2002ir}
\begin{equation}
\mathcal{L}_{\rm eff} = \frac{\lambda '}{M} \partial_\mu \phi J^\mu,
\label{eff_Lag1}
\end{equation}
where $\lambda '$ is the coupling constant, $J^\mu$ is the non-conserved baryon current and $M$ is the cut-off scale to be fixed from observation.
Assuming that the field $\phi$ is homogeneous as we work in the FRW background, we have
\begin{eqnarray}
\mathcal{L}_{\rm eff} = \frac{\lambda '}{M} \dot{\phi} \Delta n \equiv \mu (t) \Delta n,
\end{eqnarray}
where $\mu (t) \equiv \lambda ' \dot{\phi}/M$ is the effective time dependent chemical potential and $\Delta n \equiv J^0$ is the baryon number density corresponding to the global symmetry. 

In thermal equilibrium, we have
\begin{equation}
\Delta n (t;\xi) =\bar{g} \int \frac{d^3 \vec{p}}{(2 \pi)^3} [f(E,\mu)-f(E,-\mu)],
\end{equation}
where $\xi \equiv \mu / T$. For $\xi <1$,
in the first order approximation, we have
\begin{equation}
\Delta n (T; \xi) \simeq \frac{\bar{g} T^3}{6} \xi + \mathcal{O} (\xi^2) \simeq \frac{\lambda ' \bar{g}}{6M} T^2 \dot{\phi},
\end{equation}
with $\bar{g}$ being the number of degrees of freedom of the respective field corresponding to the baryon current.
The entropy density is given by
\begin{equation}
s = \frac{2 \pi^2}{45} g_* T^3,
\end{equation}
where $g_*$ is the relativistic degrees of freedom in thermal equilibrium at temperature $T$. 

The freeze-out value of the baryon to entropy ratio can be computed as 
\begin{equation}
\label{def:eta_F}
\eta_F \equiv \frac{\Delta n}{s} \Bigg|_{T=T_F} \simeq 0.38 \lambda ' \left( \frac{\bar{g}}{g_*} \right) \frac{\dot{\phi} (T_F)}{MT_F}.
\end{equation}
Quintessential baryogenesis is effective at temperatures $T_F<T<T_{\rm th}$ and since $T_{\rm r}< T_{\rm th}$, the interactions that violates the baryon number are in equilibrium during the kinetic regime during which the scalar field evolves as
\begin{equation}
\label{eq:phi_evol}
\dot{\phi} \simeq \sqrt{V_{\rm end}} \left( \frac{a_{\rm end}}{a} \right)^3.
\end{equation}

Further, we need to check the effects of the back-reaction due to  coupling of the field to $J^\mu$. After taking into account the effective term in the Lagrangian density, the evolution equation of $\phi$ becomes
\begin{equation}\label{eq:backReaction}
\left[ 1+ \frac{\lambda ' \bar{g}}{6} \left( \frac{T}{M} \right)^2 \right] (\ddot{\phi} + 3 H \dot{\phi}) + V'(\phi) = 0.
\end{equation}
Clearly, the extra terms in the field equation can be ignored for $T < M$ and the back reaction on the scalar field dynamics can be neglected. In order to give an estimate on the extra factor of Eq.~(\ref{eq:backReaction}), we use the Carroll bound \cite{Carroll:1998zi,Trodden:2003yn} $\lambda'\Mpl /M<8$. And this give us the the following upper bound on the back-reaction term 
\begin{eqnarray}
\label{breaction}
\frac{\lambda' \bar{g}}{6}(T/M)^2< \frac{\lambda' \bar{g}}{6}(T_{\rm end}/M)^2 \lesssim \frac{1 }{6}(8T_{\rm end}/\Mpl)^2\approx 1.23 \times10^{-9},
\end{eqnarray}
where we have used the estimate of $T_{\rm end}$ from Eq.~(\ref{eq:tend2}) and the fact that numerical values of coupling
 $\lambda'$ does not exceed unity  and $\bar{g}=\mathcal{O}(1)$ along with the approximation, $H_{\rm end}\simeq H_{\rm inf}$. Since the correction in the parenthesis in Eq.~(\ref{eq:backReaction}) is very small compared  to unity, we conclude that the back reaction is negligibly small and  can safely be ignored in the evolution equation (\ref{eq:backReaction}).

In this scenario, the kinetic regime is long and it it reasonable to assume that Freezing takes place before the commencement of radiative regime where we can estimate the evolution of $\phi$ in general setting. Thus, in general, we can constrain $\gamma\equiv  a_{F}/a_{\rm th}\lesssim 10^4$ (see Eqs.~(\ref{eq:TthF}) and (\ref{Tr})).

\begin{figure}[ht]
\centering
\includegraphics[scale=0.60]{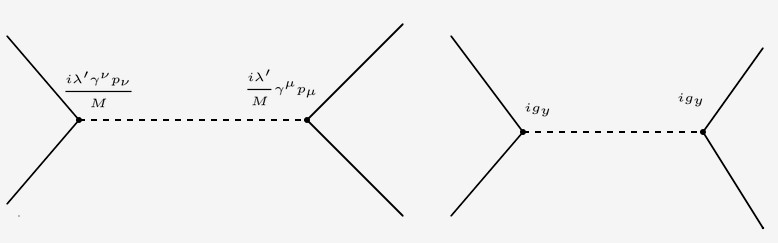}
 \caption{Feynman diagrams for left panel CPT violating interaction   given in \ref{eff_Lag1} , the non conserved baryon current is $\bar q \gamma^\mu q$, ,  in the right panel a four Fermi type interaction with Yukawa coupling mediated by the scalar field. }
\label{feyn}
\end{figure}

We have not yet fixed the concrete physical process for baryon number violation required to estimate $T_F$ that we do later. Keeping the discussion general, let us write down the expression for
the freeze-out value of the baryon to entropy ratio in the scenario of instant preheating  using Eqs.~(\ref{phid}),(\ref{Hend2}), (\ref{def:eta_F}) and (\ref{eq:phi_evol}), 
\begin{eqnarray}
\label{etaf21}
\eta_F \approx 5.4\times 10^{-3}\frac{\lambda' H_{\rm end}\Mpl}{M T_F}\left(\frac{a_{\rm end}}{a_F}\right)^3,
\end{eqnarray}
where we have taken $\frac{\bar g}{g_*}\approx10^{-2}$ and $T_F= \frac{T_{\rm th}}{\gamma}$, one gets
\begin{eqnarray}
\label{etaf22}
\eta_F &\approx& 5.4\times 10^{-3}\frac{\lambda' H_{\rm end}\Mpl}{M \gamma^2 T_{\rm th}}\left(\frac{a_{\rm end}}{a_{\rm th}}\right)^3.
\end{eqnarray}
Finally using Eqs.~(\ref{eq:tend2}) and (\ref{eq:TthF}), we find,
\begin{eqnarray}
\label{etao}
\eta_F=1.86\times10^{-2}\lambda' \frac{\alpha^2\Mpl}{M \gamma^2}.
\end{eqnarray}
It is interesting to note that the above estimate of $\eta_F$ is independent of inflationary era as $r$ drops out from Eq.~(\ref{etao}) as it should be.  Secondly, for viable numerical values of couplings $\alpha$, $\lambda'$ and the coefficient $\gamma$, making appropriate choice for the cut-off can easily land to the required baryon asymmetry, see Eq.~(\ref{etao}). If we adopt the Carroll bound $\lambda' M_{\rm Pl}/M \sim 8$, we find that $\gamma\approx 3\times10^3$ for the observed asymmetry. Let us emphasize that our estimates are independent of the underlying physical processes  to be used for baryon number violation\footnote{Needless to mention that the estimates relate to the paradigm of quintessential inflation rather than to a specific model.}. Specifying the process would provide constraint on $T_F$ or $\gamma$. Now from our formulation, we always have $T_{\rm th}\leq T_{F}\leq T_{\rm r}$. This enable us determine the upper-bound of $\gamma$ to be $10^4$ which in turns give the cut-off scale $M\approx 1.7 \times 10^{-2}\Mpl$. The minimum possible value of $\gamma$ is set for any $T_F\leq T_{\rm th}$ subject to the cut-off scale, $M$ does not exceed Planck mass. We found it  to be 1350. This is to be mentioned here that in the above consideration, we have taken $\lambda' \approx 1$ and $\alpha\approx 0.1$ for the estimation of $\gamma$; these values are sensitive to the parameters mentioned.\\
\\
{ We now estimate the amount of CPT violation near the electro-weak scale around $100 \rm GeV$ by calculating the cross section of such process . We  show  that the magnitude of CPT violating process   is much smaller than the similar kind of standard non CPT violating process at that temperature.
We estimate the cross sections of two body processes, standard (Fig.\ref{feyn})(a) and CPT violating (Fig.\ref{feyn})(b) designated by $\sigma_Y$ $\&$ $\sigma_{CPT}$,
\begin{eqnarray*}
 \sigma_Y&=& \frac{\alpha_Y^2}{T^2}\\
 \sigma_{CPT}&=& \lambda'^4 \frac{T^2}{M^4}
\end{eqnarray*}
The ratio of the two cross section for $\lambda'\approx1$ is given by around one TeV is
\begin{equation}
    \frac{\sigma_{CPT}}{\sigma_Y}\approx 10^{25} \frac{T^4 }{M^4} \simeq 10^{-30}
\end{equation}{}
where we used the bound on the cut off scale,  $M\approx 10^{-2} \Mpl$}. The CPT violating estimate is much below the laboratory bound quoted in the literature 
which is not surprising as the said process at low energies is suppressed by the cut off.
%%%%%%%%%%%%%%%%%%%%%%%%%%%%%%%%%%%%%%%%%%%%%%%%%%%%%%%%%%%%%%%%%%%
\subsection{Baryon number violation through non-renormalizable operators}
The observed baryon asymmetry in the Universe is beyond
the standard model of particle physics, thereby the effect should manifest through an effective interaction in the standard model. For instance, the theory preserves $B-L$; we could, however, imagine  an effective  four Fermi construct that would violate this symmetry.
To this effect, let us consider the following 4-fermion operator,
\begin{equation}
\label{fermiop}
\mathcal{L}_{B-L} = \frac{\tilde{g}}{M_X^2} \psi_1 \psi_2 \bar{\psi}_3 \bar{\psi}_4,
\end{equation}
where $\psi_i$ are the standard model fermions and $\tilde{g}$ is the dimensionless coupling constant which is obtained after the $B-L$ violating effects of a particle of mass $M_X$ are integrated out. The rate of processes that violates the baryon number due to this operator is defined as 
\begin{equation}
\Gamma_{B-L} = \langle \sigma (T) n(T) v \rangle,
\end{equation} 
with $\sigma (T)$ being the cross section for $\psi_1 + \psi_2 \rightarrow \psi_3 + \psi_4$, $n(T)$ is the number density of $\psi$ particles, $v$ is the relative velocity and the angular brackets $\langle ... \rangle$ denotes the thermal average.
For $T<M_X$, we have $n(T) \sim T^3$, $\sigma (T) \sim \tilde{g}^2 T^2/M_X^4$ and $v \sim 1$, therefore,
\begin{equation}
\Gamma_{B-L} (T) \simeq \frac{\tilde{g}^2}{M_X^4} T^5.
\end{equation}
As the Universe expands, the rate of interactions which are in thermal equilibrium in the early Universe drops off and they cease to be in equilibrium at the temperature $T_F$ which can be obtain from
\begin{equation}
\Gamma_{B-L} (T_F) = H(T_F).
\end{equation} 
Therefore, since during kination domination $H\propto\sqrt{\rho_\phi}\propto T^3$, we find
\begin{equation}
T_F = \frac{M_X^2}{\tilde{g}} \sqrt{\frac{H_{\rm end}}{T_{\rm end}^3}} \,,
\end{equation}
which, on using $T_F=T_{\rm th}/\gamma$, gives
\begin{equation}
    M_X=\left( \frac{T_{\rm th}^2}{\gamma^2} \tilde{g}^2\frac{T_{\rm end}^3}{H_{\rm end}}\right)^{1/4}.
\end{equation}
Now considering the instant preheating scenario for $T_{\rm th}$, from Eq.~(\ref{eq:TthF}) and $T_{\rm end}$ from Eq.~(\ref{eq:tend2}), we get
\begin{equation}
    M_X\simeq 6.5 \times 10^{-3} g^{3/4} r^{1/8} \sqrt{\frac{\tilde{g} \alpha}{\gamma}}\Mpl,
\end{equation}
where we have taken $g_*\approx 100$.  Using the BBN constraint on $g$ gives the bound on the cut-off $M_X$ for the process under consideration,
\begin{equation}
\label{MX}
    M_X\gtrsim 3.9 \times 10^{-5}\sqrt{r} \sqrt{\frac{\tilde{g} \alpha}{\gamma}}\Mpl\,,
\end{equation}
where the upper bound of $M_X$ is obtained by taking $g=1$ which is well below $M_{Pl}$.
As mentioned before, in the framework under consideration,  inflation is essentially  followed by kinetic regime during which  energy density in gravitational waves enhances compared to field energy density; nucleosynthesis  puts constraint on the efficiency of underlying reheating process. 

Using the Caroll bound in Eq.~(\ref{etao}), we must have $\gamma\approx 3\times 10^3$. We use this value to estimate the bound of $M_X$ in Eq.~(\ref{MX}) which yields
\begin{equation}
    \label{Mxb}
    M_X\gtrsim 10^{11}  \rm GeV\,,
\end{equation}
for generic values of the couplings and tensor-to-scalar ratio of perturbations $r$. Clearly, the estimate of $\gamma$ is consistent with our assumption that freezing takes place in the kinetic regime before radiation domination such that $\gamma \lesssim 10^{4}$. Thus the physical process under consideration can give rise to a viable value of $T_F$.
An important comment about the duration of kinetic regime is in order. Actually, the field continues in kinetic regime even after the commencement of radiative regime. Indeed, the scalar field, evolving in kinetic regime, after reaching equality, $\rho_\phi(a_r)=\rho_r(a_r)$, undershoots the background and continues in the same regime till the field freezes on its potential due to Hubble damping. Before field energy density becomes constant, we can estimate $\dot{\phi}$ required for $\eta_F$ in the same way as we did for $T_F>T_r$.
However, the estimate of $T_F$ or $\gamma$ requires the knowledge of the underlying model and can be determined numerically, but clearly,  $T_F<T_r$ is admissible. 

We could also use dimension-5 lepton number violating operator dubbed Weinberg operator\cite{Weinberg:1979sa,Cai:2017mow,Ma:2006fn},
\begin{equation}
    \mathcal{L}=\frac{\lambda}{\Lambda}(LH)(LH)
\end{equation}
with L, H being standard model left handed lepton and Higgs doublets respectively. The effective interaction preserves (B-L) giving rise to baryon asymmetry through lepton number violation.
We might also think to generate asymmetry using the standard model anomaly  in Eq.(\ref{eff_Lag1}) as well,
\begin{equation}
    \partial _\mu J^\mu_B=\partial _\mu J^\mu_L=N_F\left( \frac{g^{2}}{32 \pi^2}F^{a\mu \nu}\tilde{F}_{\mu \nu}- \frac{g^{'2}}{32 \pi^2}f^{a\mu \nu}\tilde{f}_{\mu \nu}\right)
    \label{an}
\end{equation}
where g and g' are $SU(2)_L$ and $U(1)_Y$ couplings,  $\tilde{F}_{a\mu \nu}$ $\&$  $\tilde{f}_{\mu \nu}$ are the duals corresponding to abelian and non-abelian field strengths and $N_F$ is the number of fermion families. Interestingly, the standard model
numbers reconcile with the Carroll bound. The anomaly term in (\ref{an}) is not suitable to $B-L$ which is preserved in standard model.

%%%%%%%%%%%%%%%%%%%%%%%%%%%%%%%%%%%%%%%%%%%%%%%%%%%%%%%%%%%%%%%%%%%
\subsection{Baryon number violation through Yukawa type of coupling}
Let us consider a most general scenario where a heavy boson, $X$, produces baryon asymmetry via decay. The decay rate of $X$ ($T\lesssim M_X$) is given by \cite{Kolb:1990vq,Olive:1994xw} 
\begin{eqnarray}
\Gamma_D\simeq \alpha_Y M_X,
\end{eqnarray}
where $\alpha_Y$ is a Yukawa coupling, $\alpha_Y\sim \frac{1}{4\pi}(\frac{m_f}{v})^2$ with $v\approx246 ~\rm GeV$.
Now if we consider the scenario where the inflaton field is in the kinetic regime when the freeze out occurs, we have $H\propto T^3$. 
The freeze out value of the temperature is found by the usual condition $\Gamma_D\simeq H\Bigr|_{M_X=T=T_F}$ giving rise to,
\begin{equation}
    T_F\approx  \sqrt{\alpha_Y}\frac{T_{\rm end}^{3/2}}{\sqrt{H_{\rm end}}}\approx 4.6\times 10^{-3} g^{3/4}\sqrt{\alpha_Y} r^{1/8} \Mpl.
\end{equation}
In the last step, we used Eqs.~(\ref{Hinfval}), (\ref{eq:tend2}) and $g_*=100$. For a generic value of $g$ and $r$, we find $T_F\approx 10^6\rm GeV$ for coupling to light fermion making it lower than $T_r$ with $\gamma\approx10^6$, which is not the case we consider.  For the higher ferion mass, we get $T_F\approx 10^9\rm GeV$ with $\gamma\approx 10^3$ just within kinetic regime. In other words, we can say that for freeze out within kinetic regime, the decay of heavy scalar boson to heavy fermions is more favourable over its decay to lighter ones. 

\begin{figure}
\subfigure[]{
\includegraphics[scale=.9,height=5.5 cm]{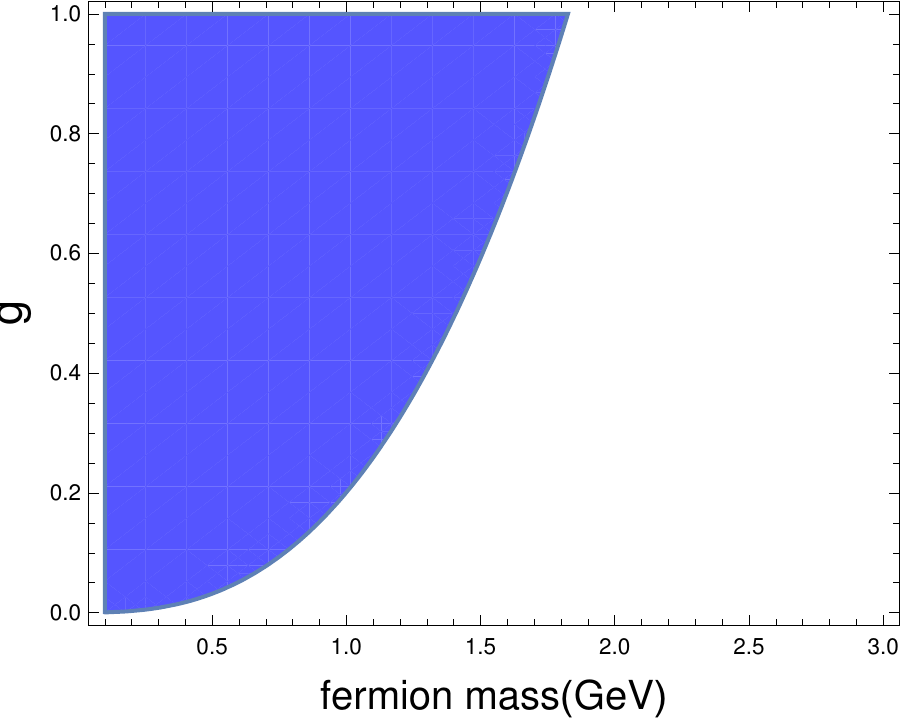}
 \label{gamma1}}
  \hspace{.8 cm} 
\subfigure[]{
 \includegraphics[width=7 cm,height=5.5 cm]{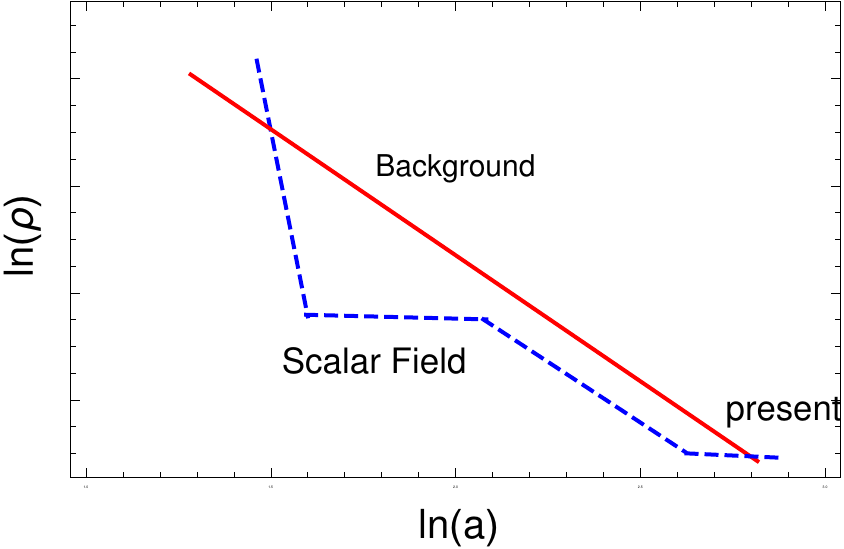}\label{gamma2}
\label{fig:test}}
\caption{Figure (a) shows the allowed region of fermion masses and the instant preheating coupling g. Lower mass  fermion couplings give rise to $\gamma\approx 10^6$ with lowest possible value of $g\gtrsim 10^{-4}$ for $r=0.05$. Higher mass fermion couplings are constrained by the lower bound on $\gamma$ necessary to keep cut-off scale $M$ below  $\Mpl$, see Eq.(\ref{etao}).
Figure (b) is a schematic diagram which shows the post inflationary evolution of $\rho_\phi$, it  depicts energy density versus the scale factor on logarithmic scale; solid red line is back ground energy density(radiation/matter) and blue dashed line depicts field energy density.}.
\end{figure}

Let us comment on the possibility of $T_F\leq T_r$. In this case, using $H \propto T^2$, we find
\begin{eqnarray}
T_F\approx \sqrt{8\pi}\alpha_Y g_*^{-1/2} M_{Pl}.
\end{eqnarray}
 For $g=1.05 \times 10^{-3} r^{1/2}$, $r=0.05$, $T_F\approx 10^6$GeV, which translates into $\gamma\approx 10^6$. Heavier fermion masses correspond to higher values of $g$ consistent with lower bound on the cut-off scale, see Fig.~\ref{gamma1}. 
As mentioned before, we can follow field evolution analytically in the kinetic regime till radiation domination. Field continues kinetic evolution thereafter also till it freezes on its potential due to Hubble damping; how long it goes on (below $T_r$) depends upon the details of the model \textit{\`{a} la} the field potential. In this case, numerical treatment is essential for estimation of $\dot{\phi}(T_F)$ included in the expression of $\eta_F $.

\subsection{Sphalaron wash out}
As demonstrated in the preceding discussion, the asymmetry in our case is generated at high temperatures around $10^{12}$ GeV. It is important to verify that the asymmetry generated survives down to electro-weak interaction and is not effected by the sphalaron wash out.
Indeed, sphalaron preserves $B-L$ but erases $B+L$\cite{Riotto:1998bt, Fukugita:1986hr},
\begin{equation}
n(t)_{B+L}\sim e^{-\frac{13}{2}N_F \frac{\Gamma_{sp}}{T^3}t}
\end{equation}
where $\Gamma_{sp}$ is the spalaron rate per unit volume above electro-weak scale, $\Gamma_{sp}\sim \alpha^4_W T^4$ and $T$. The sphalaron is effective provided that $\Gamma_{sp}/T^3\gtrsim H \to T\lesssim \alpha^4_W M_{Pl}/g_*^{1/2}\sim 10^{12} GeV$ ($T_{EW}\lesssim T \lesssim 10^{12} GeV$ ) which tells us that $B+L$ is washed out exponentially by the sphalaron above the electro-weak scale\footnote{We note that freezing temperature in our case can vary from $T_{th} \approx 10^{12} $GeV down to electro-weak scale; however, for $T_F <T_r$, one requires numerical estimates.}. This has important implication for asymmetry generated in our model. Indeed, we can write baryon number evolution as,
\begin{equation}
B(t)=\frac{1}{2}(B-L)_{t_{in}}+\frac{1}{2}(B+L)_{t_{in}}e^{-\frac{13}{2}N_F \frac{\Gamma_{sp}}{T^3}t}  
\label{master}
\end{equation}
where $t_{in}$ refers to the epoch where asymmetry freezes.  
The second term on the right hand side of (\ref{master}) practically vanishes with a time scale of the order of $ 2 T^3/13 N_F \Gamma_{sp}$ and in particular case of $L=0$, we have,
\begin{equation}
    B(T_{EW}) \approx B(T_F)
\end{equation}
and asymmetry  survives down to the electo-weak scale.
Below the elrctro-weak scale, the sphalaron rate $\Gamma_{sp}$ is suppressed by sphalaron energy and has no impact on the asymmetry thereafter. 
In what follows, we bring out constraints on the radiation density due to relic gravity waves produced during inflation and study prospects of their detection in the proposed observational missions.

%%%%%%%%%%%%%%%%%%%%%%%%%%%%%%%%%%%%%%%%%%%%%%%%%%%%%%%%%%%%%%%%%%%
\section{Relic gravitational wave constraint}
%%%%%
The detection of relic gravitational waves would be the cleanest signal for falsification of the paradigm of inflation. As for quintessential inflation, it is necessarily followed by kinetic regime which induces blue tilt in the spectrum of relic gravitational waves irrespective of the underlying model \cite{Giovannini:1998bp,Giovannini:1999bh,Giovannini:1999qj,Riazuelo:2000fc,Sahni:2001qp,Tashiro:2003qp,Giovannini:2008tm,Artymowski:2017pua,Figueroa:2018twl}. Thus the spectrum of relic gravitational waves can clearly distinguish the quintessential inflation from the standard scenario. The forthcoming gravitational wave missions would be crucial for falsifying the paradigm of inflation. In what follows, we check the relic gravitational waves spectrum against the sensitivities of proposed observational missions. It should be mentioned that our estimates presented here are also model independent. 

Gravitational wave is described as a transverse-traceless part of the metric perturbation in a spatially flat FLRW background, such like $ds^2 =-dt^2+a^2(t)(\delta_{ij}+h_{ij})dx^idx^j$. The tensor perturbations $h_{ij}$ satisfy the transverse-traceless conditions, $h_{00}=h_{0i}=\partial^ih_{ij}=h^i_i=0$, and we decompose it into its Fourier modes with the comoving wave vector $\textbf{k}$ as,
\begin{equation}
h_{ij}(t,\textbf{x})=\sum_{\lambda=+,\times}^{}\int\frac{d^3k}{(2\pi)^{3/2}}\epsilon_{ij}^{\lambda}
(\textbf{k})h_\textbf{k}^{\lambda}(t,\textbf{k})e^{i\textbf{k}\cdot\textbf{x}},
\end{equation}
where the polarization tensors $\epsilon_{ij}^{+,\times}$ satisfies
symmetric and transverse-traceless condition and are normalized as
$\sum_{i,j}^{}\epsilon_{ij}^{\lambda}(\epsilon_{ij}^{\lambda^{\prime}})^*=2\delta^{\lambda\lambda^{\prime}}$. Then the equation for gravitational waves is given by
\begin{equation}
h^{\lambda ''}_\textbf{k} (\tau) + 2 \frac{a'}{a} h^{\lambda '}_\textbf{k} (\tau) + k^2 h_\textbf{k}^{\lambda} (\tau) = 0,
\label{eq:evoh}
\end{equation}
where $\tau$ is the conformal time defined by $d\tau\equiv d t/a(t)$ and prime denotes derivative with respect to it.  The energy spectrum of gravitational waves is defined as
\begin{equation}
\Omega_{\rm GW} (k, \tau) \equiv \frac{1}{\rho_{\rm crit} (\tau)} 
\frac{d \rho_{\rm GW}}{ d \ln k },
\end{equation}
where $\rho_{\rm crit} \equiv \frac{3 H^2 (\tau)}{8 \pi G}$ and the gravitational energy density $\rho_{\rm GW}$ is given by
\begin{eqnarray}
\rho_{\rm GW} = - T^{0}_{0} 
&=& \frac{1}{64 \pi G} 
\frac{{\left( h'_{ij} \right)}^2 + {\left( \nabla h_{ij}\right)}^2}{a^2},\nonumber \\
& = & \frac{1}{32 \pi G} \int \frac{d^3 k}{(2\pi)^3}\frac{k^2}{a^2} 2 \sum_\lambda|h_k^\lambda|^2.
\end{eqnarray}
Then the fractional energy density of relic gravitational waves today can be written as,
\begin{equation}
\label{rhog}
\Omega_{\rm GW,0}=\frac{1}{12}\left(\frac{k^2}{a_0^2 H_0^2}\right)\Delta_{T,{\rm prim}}^2(k) T^2(k),
\end{equation}
where we have divided the power spectrum into primordial spectrum and transfer function as $\Delta_{T}^2(k) \equiv \frac{k^3}{\pi^2} \sum_\lambda|h_\textbf{k}^{\lambda}|^2 = \Delta_{T,{\rm prim}}^2(k) T^2(k)$.  The primordial spectrum $\Delta_{T,{\rm prim}}^2(k)$ is determined by the Hubble rate during inflation as \cite{Baumann:2009ds}
\begin{equation}
\Delta_{T,{\rm prim}}^2(k)
= \frac{2}{\pi^2}\frac{H_{\rm inf}^2}{M_{\rm Pl}^2}\bigg|_{k=aH}.
\end{equation}
The transfer function $T^2(k)$ describes the decay of GW amplitude after inflation. From Eq. \eqref{eq:evoh}, we find the amplitude stays constant when $a^\prime/a\gg k$ and starts to decay as $h_\textbf{k}^{\lambda}\propto a^{-1}$ after the mode enter the horizon ($a^\prime/a\ll k$). Thus, the transfer function is given by \cite{Kuroyanagi:2008ye}
\begin{equation}
T^2(k) = \frac{1}{2}\frac{a_{\rm hc}^2}{a_0^2},
\end{equation}
where ``hc'' denotes the value at the horizon crossing after inflation and ``0'' denotes the value today, and we set $a_0=1$.  Note that because of the factor $1/2$ in the transfer function, the value of $\Omega_{\rm GW,0}$ calculated in this paper is the root-mean-square of the GW oscillation amplitude, not the oscillation peak amplitude.  Using the Friedmann equation, the Hubble expansion rate can be expressed in terms of the density parameters as\footnote{As mentioned before, the minimum in the run away potential, specific to quintessential inflation, may be triggered at late times by coupling of scalar field to massive neutrino matter;  $\Omega_{\Lambda0}$ is then given by the minimum of the effective potential, see Ref. \cite{Hossain:2014zma} for details. }
\begin{equation}
H\approx  H_0\sqrt{
\Omega_\phi(a)+
\left(\frac{g_*}{g_{*0}}\right)
\left(\frac{g_{*s}}{g_{*s0}}\right)^{-4/3} 
\Omega_{\rm r0}\left(\frac{a}{a_0}\right)^{-4}
+\Omega_{\rm m0}\left(\frac{a}{a_0}\right)^{-3}
+\Omega_{\rm \Lambda 0}},
\label{eq:Hubble}
\end{equation}
and the contribution from the scalar field (whose dynamical contribution can be incorporated analytically only during kinetic domination) can be written as\footnote{Though Eq.~(\ref{eq:Hubble}) records the contribution of scalar field energy density during kinetic regime, i.e., for $a_{\rm end} \leq a\leq a_r$, it is approximately correct in general. Indeed, in the framework of quintessential inflation, field remains in the kinetic regime for long time. After radiation dominance, scalar field energy density remains sub-dominant throughout the history of Universe; only around the present epoch, field shows up giving rise to late time acceleration (see Fig.\ref{fig:test}) such that its dimensionless energy density parameter is well approximated by
$\Omega_{\Lambda 0}$. Thus the cosmological dynamics with (\ref{eq:Hubble}) mimics evolution of scalar field in the presence of background energy density (radiation $\&$ matter) in the framework  under consideration.}
\begin{eqnarray}
\Omega_\phi(a)&\equiv&\frac{8\pi G}{3H_0^2}\,\rho_{\phi,\rm end}\left(\frac{a_{\rm end}}{a}\right)^{\! 6}
=\frac{8\pi G}{3H_0^2}\frac32\,V_{\rm end}\left(\frac{a_{\rm end}}{a}\right)^{\! 6}
=\frac{8\pi G}{3H_0^2}\,3M_{\rm Pl}^2 H_{\rm end}^2\left(\frac{a_{\rm end}}{a}\right)^{\! 6}\nonumber\\
&=&\frac{H_{\rm end}^2}{H_0^2}\left(\frac{a_{\rm end}}{a}\right)^{\! 6}
=\frac{H_{\rm end}^2}{H_0^2}\left(\frac{a_{\rm end}}{a_r}\right)^{\! 6}\left(\frac{a_r}{a_0}\right)^{\! 6}\left(\frac{a_0}{a}\right)^{\! 6}
=\frac{H_{\rm end}^2}{H_0^2}\left(\frac{a_{\rm end}}{a_r}\right)^{\! 6}\left(\frac{T_0}{T_r}\right)^{\! 6}\left(\frac{a_0}{a}\right)^{\! 6},
\end{eqnarray}
where
$H_{\rm end}=1.4\times10^{-4}\,r^{1/2}\,M_{\rm Pl}$
and $a_{\rm end}/a_r$ and $T_r$ can be found on using Eqs.\ (\ref{eq:arbaend}) and (\ref{Tr}), respectively.
In Eq.~(\ref{eq:Hubble}), $g_*$ and $g_{*s}$ are the effective number of relativistic degrees of freedom for radiation energy density and  entropy density, respectively, and their present values are given by $g_{*0}=3.36$ and $g_{*s0}=3.91$ \cite{Kolb:1990vq}. For cosmological parameters, we use $h=0.674$, $\Omega_{\rm r0}h^2=4.15\times 10^{-5}$, $\Omega_{\rm m0}=0.315$ and $\Omega_{\rm \Lambda 0}=1-\Omega_{\rm m0}$ \cite{Aghanim:2018eyx}. Using Eq.~\eqref{eq:Hubble} and $k=a_{\rm hc}H_{\rm hc}$, the scale factor $a_{\rm hc}$ for matter and radiative regimes can be written respectively as  
\begin{eqnarray}
a_{\rm hc, MD} &=& \frac{a_0^3 H_0^2}{k^2}\Omega_{\rm m0},
\label{eq:aMD}\\
a_{\rm hc, RD} &=& \frac{a_0^2 H_0}{k}\sqrt{\Omega_{\rm r0}}
\left(\frac{g_*}{g_{*0}}\right)^{1/2}
\left(\frac{g_{*s}}{g_{*s0}}\right)^{-2/3}.
\label{eq:aRD}
\end{eqnarray}
The scale factor during the kinetic regime is given using the mode which enters at the commencement of the radiative regime $k_r=a_r H_r$ as 
\begin{equation}
\frac{a_{\rm hc, KD}}{a_r}
= \left(\frac{k_r}{k}\right)^{1/2}.
\label{eq:aKD}
\end{equation}
Substituting Eqs.\ \eqref{eq:aMD}, \eqref{eq:aRD}, and \eqref{eq:aKD} to the transfer function, today's GW amplitudes which entered the horizon during the matter, radiative, and kinetic regimes are given respectively by 
\begin{eqnarray}
\Omega_{\rm GW, 0}^{\rm (MD)}
&=& \frac{1}{6\pi^2}\Omega_{\rm m0}^2\frac{a_0^2 H_0^2}{k^2}
\frac{H_{\rm inf}^2}{M_{\rm Pl}^2}
~~~~~~~~~~  
(k_{\rm h}<k \leq k_{\rm eq}),
\\
\Omega_{\rm GW, 0}^{\rm (RD)}
&=& \frac{1}{6\pi^2}\Omega_{\rm r0}\frac{H_{\rm inf}^2}{M_{\rm Pl}^2}
\left(\frac{g_*}{g_{*0}}\right)
\left(\frac{g_{*s}}{g_{*s0}}\right)^{-4/3}
~~~~~~~~~~  
(k_{\rm eq}<k \leq k_{\rm r}),
\label{eq:OGWRD}
\\
\Omega_{\rm GW, 0}^{\rm (KD)}
&=& \Omega_{\rm GW, 0}^{\rm (RD)}\left(\frac{k}{k_r}\right)
~~~~~~~~~~  
(k_{\rm r}<k \leq k_{\rm end}),
\label{eq:OGWKD}
\end{eqnarray}
where $k_{\rm h}$, $k_{\rm eq}$, $k_{\rm r}$, and $k_{\rm end}$ represent the modes which enter the horizon at the present, radiation-matter equality, commencement of radiative regime, and end of inflation, respectively\footnote{We keep in mind that kinetic regime commences soon after inflation ends ($a_{\rm kin}\approx a_{\rm end}$).}. The transition frequencies $f=k/(2\pi)$ are given as 
\begin{eqnarray}
f_{\rm h} &=& \frac{a_0 H _0}{2\pi} 
\sim 3.5 \times 10^{-19}\mbox{Hz}\, , \\
f_{\rm eq} &=& \frac{a_{\rm eq} H _{\rm eq}}{2\pi} 
= \frac{\sqrt{2}H_0}{2\pi}\frac{\Omega_{\rm m}}{\sqrt{\Omega_{\rm r}}}
\sim 1.6 \times10^{-17}\mbox{Hz} \, , \\
f_r &=& \frac{a_r H _r}{2\pi} 
= \frac{\left(\frac{43}{11}\right)^{1/3} g_*^{1/6} T_0 T_r}{6\sqrt{10}M_{\rm Pl}}
\sim 1.6\times 10^3 \left(\frac{T_r}{10^{10}{\rm GeV}}\right)  \mbox{Hz}\, ,\\
f_{\rm end} &=& \frac{a_{\rm end} H _{\rm end}}{2\pi} 
= \frac{\left(\frac{43}{11}\right)^{1/3}
\left(\frac{\pi}{5}\right)^{1/4}
H_{\rm end}^{1/2}T_0}
{\sqrt{6}g_*^{1/12}g^{1/2}M_{\rm Pl}^{1/2}}
\sim 8.9 \times 10^8
\left(\frac{1}{g}\right)^{1/2}
\left(\frac{H_{\rm inf}}{10^{14}{\rm GeV}}\right)^{1/2}
\mbox{Hz}\, .
\label{eq:fend}
\end{eqnarray}
Here, for the temperature of the Universe before the neutrino decoupling, we have used 
\begin{equation}
    \frac{a}{a_0} = \left(\frac{11}{43}g_*\right)^{-1/3}\frac{T_0}{T},
\end{equation}
which can be obtained from the entropy conservation \cite{Kuroyanagi:2014qaa}, and we have substituted $T_0=2.73$K and $g_*=106.75$ for $T>1$ MeV.
The dominant contribution to energy of relic gravitational waves comes from the transition from inflation to kinetic regime \cite{Sahni:2001qp}.  By using Eq. \eqref{eq:aKD} and substituting Eqs.\ \eqref{eq:arbaend} and \eqref{eq:OGWRD} to Eq. \eqref{eq:OGWKD}, we find that today's amplitude of the GW which entered the horizon at the transition ($a=a_{\rm kin}\approx a_{\rm end}$) is given by 
\begin{eqnarray}
\label{BBNeq}
\Omega_{\rm GW, 0}^{\rm (peak)}
= \Omega_{\rm GW,0}^{\rm (RD)}\left(\frac{a_r}{a_{\rm end}}\right)^2
= \frac{2\pi}{g^2}\Omega_{\rm r0}\frac{H_{\rm inf}^2}{M_{\rm Pl}^2}
\left(\frac{g_*}{g_{*0}}\right)
\left(\frac{g_{s*}}{g_{*s0}}\right)^{-4/3}.
\end{eqnarray}
By combining this equation and the BBN constraint \cite{Figueroa:2018twl,Cyburt:2015mya},
\begin{equation}
\label{BBNB}
\Omega_{\rm GW, 0}h^2<1.12\times 10^{-6},
\end{equation}
we obtain the lower bound on the coupling constant $g$,
\begin{equation}
g > 1.05\times 10^{-3}\sqrt{r}.
\label{eq:glimit}
\end{equation}

Let us bring back the remark on the gravitational particle production mentioned earlier. The mechanism is such that the particle production takes place due to non-adiabatic change in the space time geometry during inflationary to non-inflationary transition. Unfortunately, the process is inefficient and might be challenged by the nucleosynthesis constraint. Indeed, the radiation energy density produced by gravitational particle production is given by,
\begin{equation}
\label{gwp}
\rho_r\simeq 0.01\times g_*H^4_{\rm end}.
\end{equation}
Using Eq.~(\ref{gwp}) and the relation, $(a_r/a_{\rm end})^2=(\rho_\phi/\rho_r)_{\rm end}$, we find \footnote{We used the approximation, $H_{\rm inf} \approx H_{\rm end}$.},
\begin{equation}
 \label{bbnf}
  \Omega^{({\rm peak})}_{{\rm GW},0}
  = \Omega_{\rm GW,0}^{\rm (RD)}\left(\frac{a_r}{a_{\rm end}}\right)^2
  \simeq\frac{10^2}{2 \pi^2g_{*,0}}\Omega_{r0}\left(\frac{g_{*,s}}{g_{*,s0}} \right) ^{-4/3}.
\end{equation}
Eq.(\ref{bbnf}) implies that,
\begin{equation}
   \Omega^{({\rm peak})}_{{\rm GW},0}h^2 \simeq 0.8 \times 10^{-6}, 
\end{equation}
which is close to  the numerical value, the BBN bound saturates at (see Eq.~(\ref{BBNB})); the latter is related to smaller numerical value of radiation density produced in gravitational particle production. The latter justifies the use of instant preheating which is quite efficient and can easily reconcile with the nucleosynthesis constraint.
In the discussion to follow, we briefly describe the future missions on gravitational waves and the prospects of their detection, in particular, the distinguished featured of the spectrum generated during quintessential inflation.

The ground-based detector network of Advanced-LIGO \cite{LIGOScientific:2019vic}, Advanced-Virgo \cite{TheVirgo:2014hva} and KAGRA \cite{Akutsu:2018axf} can reach the gravitational wave  background up to $\Omega_{\rm GW} \sim 10^{-9}$ at the frequency of $\sim 10^2$Hz when their design sensitivities are achieved. In future, satellite detectors such as Laser Interferometer Space Antenna (LISA) \cite{Audley:2017drz} and Deci-Hertz Interferometer Gravitational-wave Observatory (DECIGO) \cite{Kawamura:2011zz} probe lower frequencies ($\sim 10^{-3}$Hz and $\sim 10^{-1}$Hz, respectively) with better sensitivities. In addition, the world's largest radio telescope, Square Kilometer Array (SKA) \cite{Janssen:2014dka}, will be able to access gravitational waves around $10^{-8}$Hz using a pulsar timing array.

\begin{figure}[ht]
\centering
\includegraphics[scale=0.8]{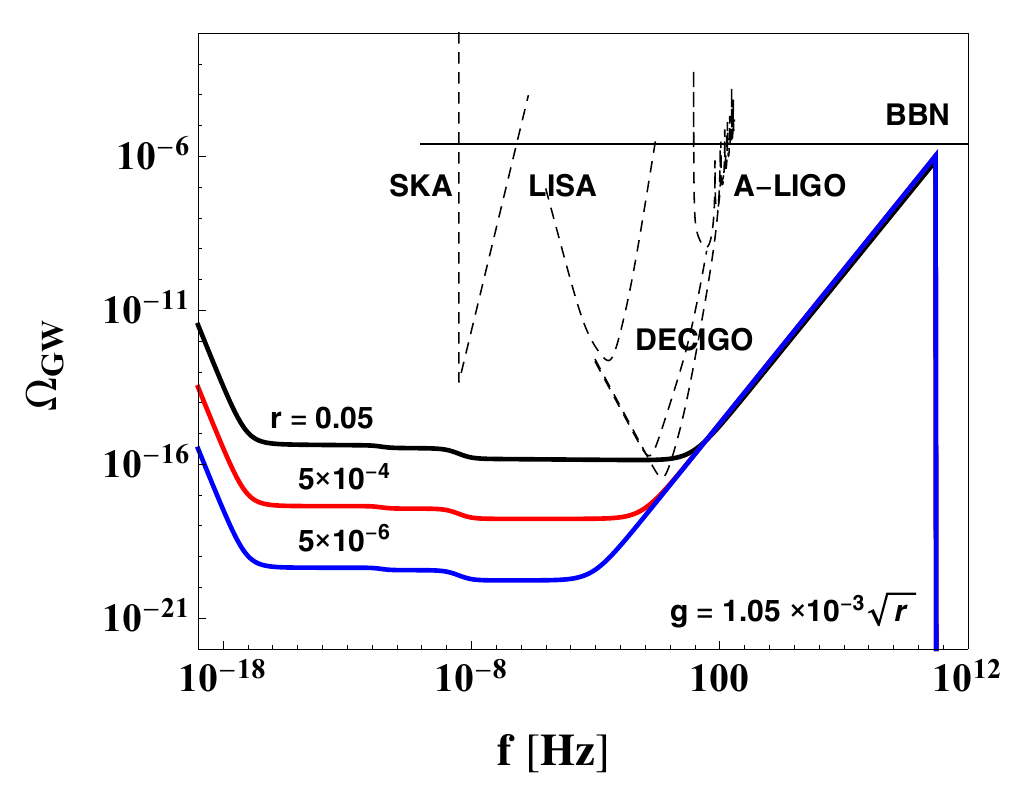}
\includegraphics[scale=0.8]{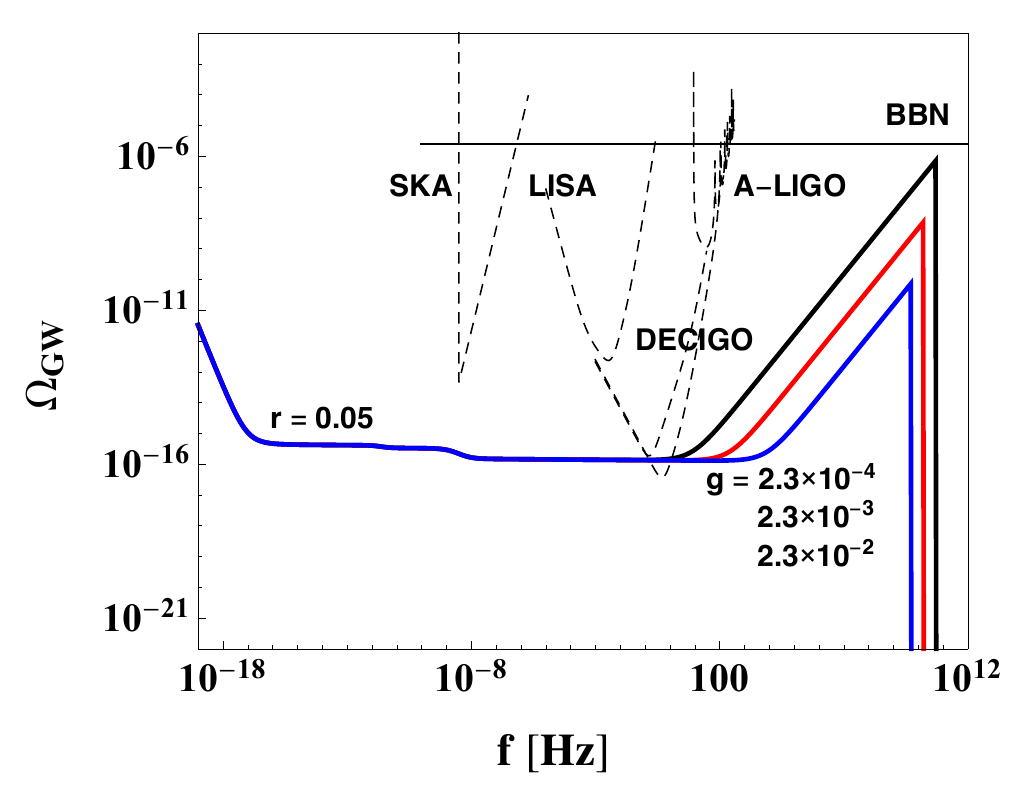}
\caption{Figure displays the relic gravitational wave spectrum; left panel corresponds to three different values of tensor-to-scalar ratio $r$ with the coupling constant $g$ being taken to the lowest value allowed by the BBN bound.  The right panel  displays the spectrum for three different values of the coupling $g$ with $r=0.05$. Dashed lines are the sensitivity curves of future GW experiments: SKA, LISA, Advanced-LIGO and DECIGO (2 lines show the original and upgraded sensitivity curves \cite{Kuroyanagi:2014qza}).
} \label{Fig:GW1}
\end{figure}
In Fig.~\ref{Fig:GW1}, we show spectra of the gravitational wave background.  In the left panel, we show the cases with $r=0.05$, $5\times 10^{-4}$, and $5\times 10^{-6}$, with the coupling constant being taken at the lower bound from the BBN limit, $g=1.05\times 10^{-3} \sqrt{r}$.  We see the blue-tilted gravitational wave spectrum at high frequencies and it peaks at $f=f_{\rm end}$, which scales as $f_{\rm end} \propto H_{\rm inf}^{1/2}/g^{1/2}$ as found from Eq.~\eqref{eq:fend}. Since $H_{\rm inf}\propto r^{1/2}$ and now we take $g\propto r^{1/2}$, the peak frequency $f_{\rm end}$ in the left panel is the same for all cases, and is too high to be tested by any types of gravitational wave experiments. In the right panel, we show the case where the tensor-to-scalar ratio is fixed at $r=0.05$ and the coupling constant is taken at the lower bound $g=1.05\times 10^{-3} \sqrt{0.05} \sim 2.3\times 10^{-4}$, shown with the cases of larger coupling constant; $2.3\times 10^{-3}$ and $2.3\times 10^{-4}$. For larger coupling constant, we see the peak frequency $f_{\rm end}\propto g^{-1/2}$ get lower, while the transition frequency $f_{\rm r}\propto T_{\rm r}\propto g^{3/2}$ shifts to higher frequencies and the peak amplitude $\Omega_{\rm GW,0}^{\rm (peak)}\propto g^{-2}$ decreases. To find parameter space of detectable gravitational waves, one may consider to lower the peak frequency by reducing the value of $H_{\rm inf}/g$. However, since the peak amplitude decreases as $\Omega_{\rm GW,0}^{\rm (peak)}\propto H_{\rm inf}^2/g^2$, the gravitational wave amplitude becomes far below the sensitivities of the experiments. In conclusion, it is difficult to probe the blue-tilted spectrum induced by the existence of kinetic regime by currently planned gravitational wave experiment. However, if we can increase the sensitivity of experiments which explores higher frequencies \cite{SRI,Cruise:2006zt,Cruise:2012zz,Arvanitaki:2012cn,Sabin:2014bua,Goryachev:2014yra,Chou:2016hbb,Robbins:2018thb,Ito:2019wcb}, one may be able to reach the signature of kinetic regime.

\section{Conclusions and future outlook}
In this paper, we investigated the possibility  of  generation of observed baryon asymmetry in the framework of quintessential inflation. We focused on the model independent features of the paradigm. For instance, irrespective of the underlying model, inflation is followed by kinetic regime in the scenario of quintessential inflation. Numerical investigation shows that the kinetic regime sets in fast after inflation ends such that $H_{\rm end}\simeq H_{\rm kin}$ is a good approximation. Inflation scale $H_{\rm inf}$ depends upon the tensor-to-scalar  ratio of perturbations $r$ and does not require the knowledge of underlying  model of inflation but its relation to $H_{\rm end}$ does. However, since tensor-to-scalar ratio of perturbations $r$ is small, to a good approximation, one can take $H_{\rm end}\simeq H_{\rm inf}$; investigation of concrete models reveal that their ratio is $\mathcal{O}(1)$. 

Since the field potential in the scenario under consideration is typically run away type, one requires an alternative mechanism for reheating in this case. One might use the gravitational particle production which is a universal process. In this case, the reheating temperature is determined by the estimate of $H_{\rm end}$. This process, however, is inefficient and we do not use it here, see Eq.\ (\ref{bbnf}) and Ref.\ (\cite{Sahni:2001qp}) for details.
%One might think to fix $T_{\rm end}$  without assuming a particular reheating mechanism, using the fact that cosmological scales which leave the horizon during inflation and reenter it later. 
%One might then arrive at a  definite relation such that the knowledge of $H_{\rm inf}, H_0, T_0$ and the number of e-foldings $\mathcal{N}$ suffice for the estimation of $T_{\rm end}$ though the latter is exponentially sensitive to $\mathcal{N}$ and can not be trusted\footnote{Indeed, one uses the mentioned relation to estimate the number of efoldings from reheating temperature given by a suitable reheating mechanism.}. In fact, a reheating mechanism is required for the knowledge of $T_{\rm end}$ whose lower bound is dictated by the nucleosynthesis constraint imposed due to relic gravitational waves produced during inflation. 
%The universal gravitational particle production fails to meet the BBN requirement, see Eq.~(\ref{bbnf}), 
In this paper, we used the instant particle production mechanism which can easily reconcile with the nucleosynthesis requirement.
For lower bound on the coupling $g$ and the upper bound on the tensor-to-scalar ratio $r$, the estimated value of radiation temperature at the end of inflation, consistent with BBN bound, is found to be, $T_{\rm end} \approx 2.3 \times10^{13} ~\rm GeV $.

As the standard model of particle physics fails to generate the required amount of baryon asymmetry, one looks beyond it by using an effective theory. For instance, one can assume an additional term in the Lagrangian of the type $\partial_\mu \phi J^\mu_B$, suppressed by a cut-off, which can formally be attributed to spontaneous violation of CPT. The latter allows to evade the  Sakharov's conditions in the equilibrium state. 
We then computed the freezing value of the asymmetry $\eta_F$, the estimate depends upon the ratio of $T_F$ and $T_{\rm th}$ designated by $\gamma$, see Eq.~(\ref{etao}). We  checked that the back reaction of the additional construct on the field evolution is negligibly small, see Eq.~(\ref{breaction}). Using the generic numerical value of the coupling associated with instant preheating, we estimated, $T_{\rm th}\approx 10^{12}$ GeV and $T_r\approx 10^{8}$ GeV. Obviously, kinetic regime is long, it even continues after the commencement of radiation domination, details depend upon the underlying model of inflation. Since our treatment is model independent, we assumed that $T_F\gtrsim T_r$ or $\gamma \lesssim 10^{4}$; the lower bound on $\gamma$ is dictated by the requirement that $M< M_{Pl}$. Our findings are model independent in terms of the underlying model of inflation as well as the underlying physical processes responsible for baryon number violation\footnote{Though the estimates are specific to model of spontaneous baryogenesis.}. By invoking a definite process for baryon number violation, we could consistently fix $T_z$ or $\gamma$. As a concrete source of baryon asymmetry, in particular, we used four Fermi operator that explicitly includes  violation of $B-L$, absent in the standard electro-weak theory. We have shown that the observed asymmetry can be obtained by fixing the cut-off in the allowed region. We have demonstrated that the asymmetry generated in our model around $10^{12} GeV$ survives down to electro-weak scale.

One of the most important aspects of inflationary paradigm is associated with the quantum mechanical production
of relic gravitational waves whose stochastic
signature presents a challenge to future missions.
Their detection would unveil the important secret of early universe, namely, its inflationary beginning.
Further, the quintessential inflation has a distinguished feature; it is essentially followed by kinetic regime, which necessarily induces a blue tilt in the spectrum of relic gravitational waves. The amplitude of gravitational waves produced around the end of inflation enhances during kinetic regime which might challenge the nucleosynthesis constraint. The latter puts a restriction on reheating temperature or equivalently on the coupling $g$, see Eq.~(\ref{eq:glimit}).
We have discussed the prospects of detection of stochastic gravitational wave background by terrestrial and space-borne gravitational wave observatories such as SKA, LISA, A-LIGO and DECIGO.
It turns out to be difficult to probe the blue-tilted spectrum due to kinetic regime by currently planned gravitational wave experiment, but it gives a good motivation for building detectors at higher frequencies \cite{SRI,Cruise:2006zt,Cruise:2012zz,Arvanitaki:2012cn,Sabin:2014bua,Goryachev:2014yra,Chou:2016hbb,Robbins:2018thb,Ito:2019wcb}.

Last but not least, instead of instant preheating used in the framework under consideration, we could also invoke curvaton mechanism to do the needful. The curvaton field has a distinguished feature, it does not interact with inflaton and remains frozen during inflation. It would be interesting to revisit the
paradigm in the presence of curvaton preheating; it will also be of interest to repeat our analysis in the framework of warm inflation \cite{Dimow} and we defer the same to our future investigations.
It would certainly be important to investigate the generation of baryon asymmetry in a UV complete set up, say by extending the electro-weak theory with possibility of giving masses to neutrinos \textit{\`{a} la} baryogenesis through lepto-genesis \cite{bari}. In this case, baryon asymmetry of the Universe is due to lepton asymmetry generated in  decays  of heavy right handed neutrino.
The (assumed) coupling of scalar field to massive neutrino matter\footnote{We have in mind the paradigm of quintessential inflation where inflaton survives after the end of inflation. On phenomenological grounds, it is plausible to imagine a coupling of the scalar field to massive neutrino matter proportional to its trace. Such a coupling is absent as long as neutrinos are relativistic; coupling builds up dynamically at late times as neutrinos turn non-relativistic. The latter triggers a minimum in the effective potential of the field such that the minimum of the field potential is proportional to the neutrino mass in fourth power. Since the dark energy scale is close to the energy scale associated with neutrino mass, little adjustment of the proportionality constant should then allow to reconcile with the observed value of dark energy.}, proportional to the trace of its energy momentum tensor, would build dynamically at late stages allowing for late time acceleration. In this picture, energy density of dark energy would naturally get linked to the light neutrino mass, the only relevant physical scale in the late Universe. We shall undertake the said investigations in our future work. 

\section{Acknowledgements}
The authors thank the Yukawa Institute for Theoretical Physics at Kyoto University, where this work was initiated
during the workshop YITP-T-17-02 “Gravity and Cosmology 2018”.We thank V. Sahni for kindly reading the manuscript and pointing out corrections.
MS and NJ thank Romesh Kaul, S. Nojiri, Arnab Dasgupta and Rathin Adhikari  for fruitful discussions.  MS also thanks S. Nojiri  for  for invitation to KMI, University of Nagoya, facilitating his participation in the workshop YITP-T-17-02. SK is partially supported by JSPS KAKENHI No.17K14282 and Career Development Project for Researchers of Allied Universities. MS and NJ are partially supported by the Indo-Russia Project. NJ is  thankful to UGC, Govt. of India for financial support (INT/RUS/RFBR/P-315). MS is partially supported by the Ministry of Education and Science of the Republic of Kazakhstan, Grant No. 0118RK00935". Part of the work was done during the visit of MS to Eurasian National University, Nur-Sultan, Kazakhstan.
\\
\section{Appendix: Instant preheating in brief}
\label{append1}
%%%%%%%
As we pointed out that standard preheating is not applicable to models of quintessential inflation and that instant preheating provides with a viable alternative.
In this appendix, we present the basics of the instant preheating mechanism.
As mentioned before, instant particle production is suited to models with run away type of potentials specific to quintessential inflation. In this picture, one assumes that a scalar field  $\chi$ with bare mass zero  interacts with the inflaton($\phi$) as well as with the matter field ($\psi$),
\begin{equation}
\mathcal{L}_{int} = - \frac{1}{2} g^2 \phi^2 \chi^2 - h \bar{\psi} \psi \chi, \label{Lint1}
\end{equation}
where $g$ and $h$ are positive coupling constants with a restriction, $g,h<1$ such that a perturbative treatment is viable for the Lagrangian, Eq.~(\ref{Lint1}). The new fields occurring in the Lagrangian are supposed to be light during inflation.
The  effective mass of $\chi$ is given by,
$m_{\chi} = g \left| \phi \right|$.
The condition of  $\chi$ particle production after inflation is associated with  non adiabatic evolution of $m_\chi$,

\begin{equation}
\dot{m}_{\chi} \gtrsim m_{\chi}^2 \longrightarrow \dot{\phi} \gtrsim g \phi^2.
\end{equation}
To this effect, let us estimate  $\dot{\phi}_{\rm end}$  using the expression of slow roll parameter,
  $\epsilon \sim {\dot{\phi}^2}/{H^2}$ , { since $\epsilon_{end} = 1$, we find,}
\begin{eqnarray}
\label{dotphiend1}
 \dot{\phi}_{\rm end}^2 \simeq V_{\rm end} \to |\dot{\phi}_{\rm end}| \cong V_{\rm end}^{1/2}.
\end{eqnarray}
Thus, particle production, in the model commences provided that ,
\begin{equation}
\label{phiprod1}
\left| \phi \right| \lesssim | \phi_{\rm pd} | 
= \sqrt{\frac{| \dot{\phi}_{\rm end} |}{g}} 
= \sqrt{\frac{V_{\rm end}^{\frac{1}{2}}}{ g}} \longrightarrow g^2 \gtrsim M_{pl}^{-4} V_{\rm end}
\left( \phi_{\rm pd} \leq M_{pl} \right),
\end{equation}
where the subscript "pd" stands for particle production, i.e., the values of physical quantities at which particle production commences. Using Eqs.~(\ref{dotphiend1}) and (\ref{phiprod1}), we have
\begin{equation}
\frac{\left| \phi \right|}{| \dot{\phi} |}
\approx \frac{\left| \phi_{\rm pd} \right|}{| \dot{\phi}_{\rm end} |}
= g^{- \frac{1}{2}} {| \dot{\phi}_{\rm end} |}^{- \frac{1}{2}}.
\end{equation}
The estimated production time is then given by, 
\begin{align}
t_{\rm pd} &\approx \frac{\phi}{| \dot{\phi}| }
\approx  g^{- \frac{1}{2}} {| \dot{\phi}_{\rm end} |}^{- \frac{1}{2}}.
\end{align}
Use of uncertainty relation then allows to estimate the corresponding wave number,
\begin{align}
k_{\rm pd} \approx t_{\rm pd}^{-1} \approx \sqrt{g | \dot{\phi}_{\rm end} |},
\end{align}
which is then used to compute the  occupation number for $\chi$ particles ,
\begin{align}
n_k \approx e^{- \frac{\pi k^2}{k_{\rm pd}^{2}}}.
\end{align}
Equipped with the aforesaid, we can estimate energy density of produced particles. Indeed, their number density and energy density are given by,
\begin{eqnarray}
&& N_{\chi} = \frac{1}{(2 \pi)^3} \int_{0}^{\infty}{n_k d^3 k}
\simeq \frac{{\left( g | \dot{\phi}_{\rm end} | \right)}^{\frac{3}{2}}}{(2 \pi)^3},\\
&& \rho_{\chi} = N_{\chi} m_{\chi}
\simeq  \frac{{( g | \dot{\phi}_{\rm end} | )}^{\frac{3}{2}}}{(2 \pi)^3} 
 g \left| \phi_{\rm pd} \right|  \simeq \frac{g^2 V_{\rm end}}{(2 \pi)^3}.
 \label{rhochi}
\end{eqnarray}

Assuming that the $\chi$ field decays\footnote{In fact, $\chi$ can be made to decay fast, almost instantaneously, by demanding, $\Gamma_\chi>>H_{end}$ which in turn puts a restriction on the coupling $h$.} into relativistic degrees of freedom  and that thermalization takes place instantaneously, using Eq.~(\ref{rhochi}) and the expression, $H^2_{end}\simeq V_{end}/2M^2_{Pl}$ , one arrives at the final estimate,
\begin{equation}
{\left( \frac{\rho_{\phi}}{\rho_r} \right)}_{\rm end}\approx \frac{(2 \pi)^3}{g^2}, \label{ratiorhophirhor}
\end{equation}
which can comply with nucleosynthesis constraint if the coupling $g$ or equivalently radiation density $\rho_{r,end}$ is appropriately constrained, see Ref. \cite{Sami:2004xk} and references therein for details.

%%%%%%%%%%%%

\end{document}